\documentclass[twocolumn]{aastex631}

\graphicspath{{./}{Figures/}}

\usepackage{ulem}
\usepackage{CJK}
\usepackage{hyperref}
\usepackage{amsmath}
\usepackage{graphicx}
\usepackage{multirow}

\begin{document}

\begin{CJK*}{UTF8}{gbsn}
\title{The Catalog of early-type Runaway stars from LAMOST-DR8}

\correspondingauthor{Xuefei Chen; Luqian Wang}
\email{cxf@ynao.ac.cn; wangluqian@ynao.ac.cn}

\author[0000-0001-9989-9834]{Yanjun Guo (郭彦君)}
\affiliation{Yunnan Observatories, Chinese Academy of Sciences (CAS), Kunming 650216, Yunnan, China}
\affiliation{Max-Planck Institute for Astronomy, Königstuhl 17, D-69117 Heidelberg, Germany}
\affiliation{School of Astronomy and Space Science, University of Chinese Academy of Sciences, Beijing, 100049, People's Republic of China}

\author[0000-0003-4511-6800]{Luqian Wang (王璐茜)}
\affiliation{Yunnan Observatories, Chinese Academy of Sciences (CAS), Kunming 650216, Yunnan, China}

\author[0000-0002-1802-6917]{Chao Liu (刘超)}
\affiliation{Key Laboratory of Space Astronomy and Technology, National Astronomical Observatories, Chinese Academy of Sciences, Beijing 100101, Peopleʼs Republic of China;}

\author[0000-0002-3616-9268]{You Wu (吴优)}
\affiliation{Key Laboratory of Space Astronomy and Technology, National Astronomical Observatories, Chinese Academy of Sciences, Beijing 100101, Peopleʼs Republic of China;}

\author[0000-0001-9204-7778]{ZhanWen Han (韩占文)}
\affiliation{Yunnan Observatories, Chinese Academy of Sciences (CAS), Kunming 650216, Yunnan, China}
\affiliation{School of Astronomy and Space Science, University of Chinese Academy of Sciences, Beijing, 100049, People's Republic of China}

\author[0000-0001-5284-8001]{XueFei Chen (陈雪飞)}
\affiliation{Yunnan Observatories, Chinese Academy of Sciences (CAS), Kunming 650216, Yunnan, China}
\affiliation{School of Astronomy and Space Science, University of Chinese Academy of Sciences, Beijing, 100049, People's Republic of China}

\begin{abstract} 
Runaway stars are OB-type stars ejected from their birthplace with large peculiar velocities. The leading hypothesis addressed in their formation includes the supernova ejection mechanism and the dynamic ejection scenario. 
Identification of runaway populations is the first step to investigating their formation and evolution.
Here we present our work of searching for Galactic runaway candidate stars from the LAMOST Medium-Resolution Survey DR8 database. 
After studying the kinematic properties for a collection of 4,432 early-type stars, predominantly B-type stars, using the radial velocity measurements from LAMOST DR8 and astrometric solutions made by \emph{Gaia} DR3, we identified 229 runaway candidate stars.
They span a wide distribution in projected rotational velocities. 
We investigated the Galactic spatial distribution of the runaway population and noticed that most of them likely reside within the Galactic thin disk. 
Based upon analyzing the Doppler shifts of the candidate stars, we found two binary runaway candidates displaying velocity variation with estimated orbital periods of 40 and 61 days.
\end{abstract}


\section{Introduction}\label{sec:intro}
Early-type stars with spectral classification span from O-type to B-type are massive, luminous, and young, with high effective temperatures \citep{1968Morton,1973Morgan}. 
They are fascinating objects in a wide astronomy field for several reasons.
Early-type stars contribute to the universe's re-ionization and may enrich metallicity in the Galactic environment \citep{hopkins2014}.
Most early-type stars are found in binary systems, and they likely evolve to compact binary systems, such as double black holes, double neutron stars, or neutron star and black hole binaries \citep{2016GW1,2016GW2,2020GWbinary}.

Galactic OB runaway stars are characterized by either having peculiar velocities larger than the typical range of around $30-40$ km s$^{-1}$ or by being located far away from the Galactic plane. 
The first known examples are AE Aur and $\mu$ Col,
which have space velocities exceeding 100 km/s and are moving 
in opposite directions \citep{Blaauw1954}. 
Both stars probably originated in the Orion Nebula 
around two million years ago. 
\citet{1961Blaauw} described these stars as runaway stars 
and conducted a study of 19 runaways with space velocities greater than 40 km/s to investigate their group properties.
He found that the ratio of O/B runaways was 10 times greater than 
that found in normal low-velocity OB stars. 
Furthermore, none of the runaways were visual binaries 
or known spectroscopic binaries. As a result, 
he proposed that the runaways acquired their high velocity 
through the breakup of a binary system by a supernova explosion, 
a suggestion that had originally been made by \citet{1957Zwicky}. We refer to this model as a `binary supernova scenario' (BSS) for the reminder of the paper.

Since then many more high-velocity stars have been identified in kinematical studies of the OB stars (e.g. \citet{1978Grosbol,1984Karimova,1987Gies,1991Stone}). 
The estimated fraction of OB stars that are runaways ranges from 7\% to 49\% \citep{1979Stone,1991Stone} before 2000. 
In the first decade of the century, extensive research has been conducted for searching for runway stars based on astrometric measurements from \emph{Hipparcos} satellite \citep{ESA1997,2007vanLeeuwenV}.
\citet{Berger2001} investigated the kinematics of Be stars and found 24 runaway stars of 344 Galactic Be population 
using the criterion of peculiar space velocity lager than 40 km s$^{-1}$. 
\citet{deWit2005} studied peculiar space motions of 43 Galactic O-type field stars and found 22 of them are potential runaway candidates. 
\citet{2005Mdzinarishvili} compile a catalog of 2,228 OB brighter than $V=10$ mag from \emph{Hipparcos} and reported 61 OB runaway candidates. 
Later work by \citet{Tetzlaff2011} performed a comprehensive kinematic study of 7,663 young stars within 3 kpcs from the Sun and 
concluded 2,353 runaway candidates (45\% has no radial velocity measurements), representing the largest sample based on the \emph{Hipparcos} catalog.

During this period, several scenarios have been proposed for understanding the formation/origin of the runaway stars, 
and there are two popular models presently. 
One is the modern version of the BSS, 
which interprets the runaway stars as a stage in massive binary evolution.
The initially more massive star expands during core helium burning and transfers mass to the companion, and becomes a helium star at the end of mass transfer. If the helium star is massive enough, it soon explodes as a Type II supernova, and the recoil produced by the ejection of the supernova shell imparts a velocity of several tens of kilometers per second to the system. The binary probably remains bound and appears as a runaway OB star with a neutron star or black hole companion. After several million years, the secondary OB star expands and begins to transfer mass to the collapsed companion, expressing as massive X-ray binaries. 
The other is known as the dynamical ejection scenario (DES), as first proposed by \citet{Poveda1967} (see also \citet{2001Hoogerwerf}).  
In this scenario, the runaway velocities may be a result of unusual conditions during star formation or the dynamical evolution of a young cluster or association. Several authors have suggested that strong gravitational interactions are expected within a young cluster of OB stars \citep{1967Poveda,1968van,1971Allen}. 
Numerical simulations of close 3-body and binary-binary encounters show that these encounters can produce high-velocity stars \citep{1975Heggie,1983Mikkola,1984Hut}.

Our understanding of runaway stars has been improved significantly in recent years thanks to the extensive collection and the unprecedented accuracy of astrometric observations released by the \emph{Gaia} mission. 
\citet{Boubert2018} utilized the \emph{Gaia} astrometric solutions and found 40 Galactic Be runaway stars from a sample of 632 stars.
Adopting proper motions from \emph{Gaia} DR1, \citet{maiz2018} extended the search for O-type and BA-type supergiant runaway stars from \emph{Gaia} and \emph{Hipparcos} astrometric measurements and detected 76 runaway stars.
\cite{2020Schoettler} found 31 runaway and 54 walkaway candidates within 100 pc of the Orion Nebula Cluster, utilizing the \emph{Gaia} DR2 astrometry and photometry data. 
\cite{2021Bischoff} reanalyzed a sample of known runaways from \emph{Gaia} DR2 to validate their ages and discovered three new runaway stars. 
\citet{2022Bhat} utilized both \emph{Hipparcos} and \emph{Gaia} astrometry to compute the trajectories of known runaway stars and based upon age estimation, to trace their possible formation scenarios. 
Recently, additional samples of runaway stars have been identified through the \emph{Gaia} EDR3 \citep{Kobulnicky2022}, yielding 102 runaway O stars.
The sample of runaway stars is the essential input for kinematic simulations to investigate the spatial and number distribution of runaway stars. This information is vital to constrain their origin and evolutionary history. 

Many of the above studies lack RV measurements \citep{2011Tetzlaff,maiz2018,2020Schoettler,Kobulnicky2022}, restricting the identification of runaway stars solely to peculiar tangential velocity analysis. 
Alternatively, some obtain RV information through cross-matching from different literature sources and SIMBAD \citep{Boubert2018,2022Bhat}.
The limitation of inconsistency in RV measurements may introduce errors into the analysis. Therefore, spectroscopic studies with homogeneous RV measurements are necessary \citep{Kobulnicky2022,2022Bhat}. 
The recent release of a large collection of spectroscopic observations from the Large sky Area Multi-Object fiber Spectroscopic Telescope (LAMOST) has offered us an opportunity to search for new OB massive runaway stars using this extensive and homogeneous database \citep{2012CuiXiangQun,2012ZhaoGang,2012DengLiCai,2020LiuChao}. 
Recently, \citet{2021GYJfb,Guo2022} identified 9,382 early-type stars from the LAMOST DR7 database. 
The work by \citet{2021GYJslam} adopted a data-driven algorithm to measure the atmospheric parameters of these newly identified stars, such as the effective temperature ($T_{\rm eff}$), gravity ($\log{g}$), metallicity ($[M/H]$), and the projected rotational velocity ($V\sin{i}$). 
Motivated by the LAMOST radial velocity catalog of 3.8 million single-exposure spectra released by \citet{2021zhangboRV}, our goal for this work is to search for Galactic OB runaway stars using the LAMOST early-type star sample and the astrometric measurements from \emph{Gaia} DR3 database \citep{2021GaiaDr3}. 

The paper is organized as follows. 
We introduce the LAMOST data collection in Section 2. 
Section 3 describes the several quality filters applied to the sample. 
The kinematic analysis for identifying OB runaway stars and discussing our results is presented in Section 4. 
We summarize our conclusion in Section 5.

\section{The Sample}\label{sec:sample}
We adopt the sample of early-type stars identified from the LAMOST Medium Resolution Survey (MRS, $R=7,500$）database by \citet{2021GYJfb} to search for OB runaway stars. 
LAMOST is a 4.0-meter quasi-meridian reflecting Schmidt telescope located at Xinglong station of the National Astronomical Observatory, China. 
The focal surface of the telescope has a field of view of 5$^{\circ}$ and accommodates 4,000 fibers \citep{2012CuiXiangQun,2012ZhaoGang,2012DengLiCai}. 
LAMOST began a five-year MRS in 2018 October, with an observing strategy of obtaining time-domain observations for objects in the Galactic environment. 
The MRS spectra are made with blue and red cameras, i.e.
covering a wavelength range of $4950-5350$ \AA\ for the blue arm and $6300-6800$ \AA\ for the red arm \citep{2020LiuChao}. 
The eighth data release (DR8) dataset \footnote{\url{http://www.lamost.org/dr8/}} comprises 16.6 million spectra and provides 7.91 million sets of estimated stellar parameters.  

Using the technique of measuring the equivalent width of spectral line profiles of H$\alpha$, \ion{He}{1}, and \ion{Mg}{1},
\citet{2021GYJfb} identified a sample of 9,382 early-type stars from a collection of over 800,000 stars from the LAMOST MRS DR7 database.
The atmospheric parameters of those identified early-type massive stars have been derived by adopting a machine learning module, the stellar label machine ({\tt SLAM}) \citep{2021GYJslam}. 
We adopted the collection of 9,382 early-type stars as our initial sample and applied a series of quality filters to this sample. 
Detailed procedures are discussed below. 

\subsection{Preliminary sample}
Among the original adopted sample of 9,382 early-type stars, 
the spectra of these stars have a signal-to-noise ratio ($S/N$) of $\geq 40$. 
We cross-matched the published LAMOST MRS radial velocity (RV) survey\footnote{\url{https://github.com/hypergravity/paperdata}} released by \citet{2021zhangboRV} to collect the Doppler shifts of the sample stars. 
In their work, the authors developed a self-consistent method to obtain the absolute RV measurements for more than 0.8 million stars in the LAMOST MRS. 
The authors compared their measured RVs to those of {\it Gaia} DR2 RVs to obtain the RV Zero Point (RVZP) values, and such corrections were then applied to their measured RVs to determine the absolute velocity of each star without referring to standard stars. \citet{2021zhangboRV} report that the precision of measured RV values degrades significantly towards spectra with lower $S/N$ values (see their Fig.~4). The RV measurements achieve a precision of 0.80 km s$^{-1}$ for spectra with $S/N$ within the range between 50 and 100 at the red arm. We thus adopted the 
selection criteria of $S/N\ < 50$ and the LAMOST bad pixel flag of {\tt n\_bad\_pix\ }$\geq$ 100 to reject 1,371 low-quality spectra in the sample. Such selection narrows the working sample to 8,011 stars. We average all the velocity values for target stars with more than one RV measurement.

\subsection{Eliminating emission-line stars}
\begin{figure*}
\centering
\includegraphics[scale=0.9]{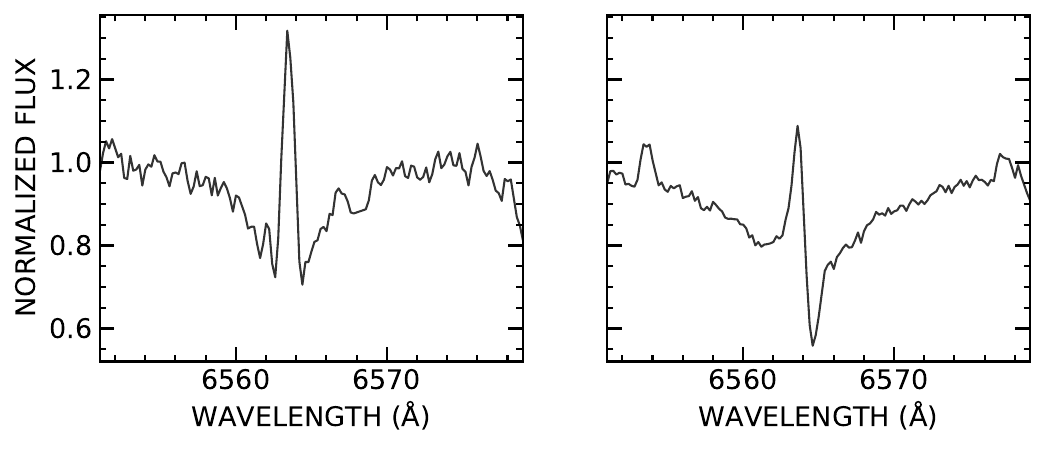}
\caption{Example spectra of H$\alpha$ emission profiles identified in the preliminary sample from the CNN {\tt ResNet} work. These spectra were eliminated from the sample due to the erroneous RV measurements reported in \citet{2021zhangboRV}. (Left panel: \emph{Gaia}~Source~ID~3437654474685782912, right panel: \emph{Gaia}~Source~ID~3326400154406915712).} \label{fig:emission-line}
\end{figure*}

\begin{figure*}
\centering
\includegraphics[scale=0.9]{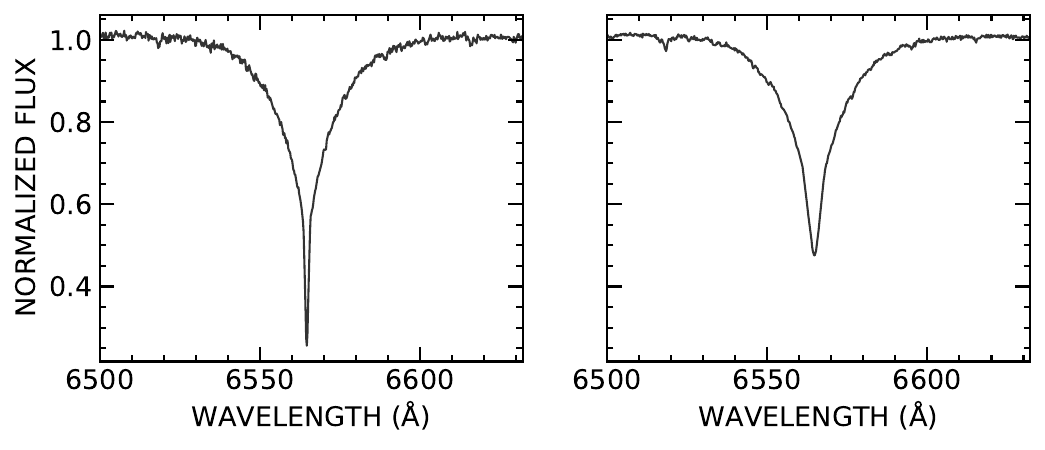}
\caption{Example spectra displaying H$\alpha$ absorption profile in our sample (Left panel: \emph{Gaia}~Source~ID~3104143533939403264, right panel: \emph{Gaia}~Source~ID~2189024507268284416).}
\label{fig:absorption-line}
\end{figure*}

In the RV catalog compiled by \citet{2021zhangboRV}, the authors noted that in the case of emission-line stars, the emission features were masked before measuring the RV. We caution that such treatment may result in erroneous RV entries in the catalog.
We thus adopted the approach reported in \citet{2022luqianBe} to eliminate the emission-line stars in our sample to ensure the accuracy of the collected RVs.

In the work of \cite{2022luqianBe}, the authors constructed a training sample of 1,042 non-emission line spectra and 1,042 emission spectra from LAMOST MRS DR7 containing spectral images at wavelengths between $6530-6590$ \AA.  
After being data augmented with geometric transformations (a regularisation method developed by \citealt{perez2017effectiveness}), these samples were fed into the {\tt ResNet} network \citep{he2016deep} for training
\footnote{{\tt ResNet} is a deep convolutional neural network (CNN) that introduces residual connections in the network to address the problem of vanishing gradients and exploding gradients, thereby enabling the construction of deeper and more accurate networks \citep{10.1007/978-3-319-46493-0_38}. 
{\tt ResNet} has greatly promoted the development and application of deep neural networks and is widely used in various computer vision tasks. }.
\citet{2022luqianBe} eventually obtained a 
spectral classifier based on a convolutional neural network that classifies spectra based on the H$\alpha$ emission features with an accuracy of 99.5$\%$. 
For a detailed description of this approach and the training process, refer to Section 4 of their paper. 

In our study, we generated spectral images, which are graphical representations of the stellar spectra containing the H$\alpha$ features, over the wavelength range of $6530-6590$ \AA. Each image was of size $256\times256$ pixels.
We then classified these spectral images using the well-trained network model reported in \citet{2022luqianBe}, and we ultimately identified 1,843 stars that exhibited emission lines. 
We excluded these emission-line stars in our sample, and visually checked the sample to further remove 182 stars displaying weak absorption profiles or featureless spectra.
Figure~\ref{fig:emission-line} shows the example spectra of H$\alpha$ emission stars excluded from the preliminary sample using the {\tt ResNet} identification task.
As a comparison, Figure~\ref{fig:absorption-line} shows the spectra of H$\alpha$ absorption profiles for target stars identified in our sample.  

\subsection{Final sample}\label{sec:final sample}
In order to identify the OB runaways stars in the cleaned sample, we cross-matched the sample stars with the \emph{Gaia} DR3 \citep{2021GaiaDr3} catalog to collect their equatorial coordinates (R.A.\ and Decl.) and the astrometric solution, including the proper motion ($\mu_\alpha$, $\mu_\beta$), the parallax ($\pi$), and their associated errors for 5,915 common stars using a circular searching window with a radius of 3\arcsec. We applied the quality filter of the renormalized unit weight error ({\tt RUWE}) $>$ 1.4 \citep{Lindegren2018} to reject 1,333 stars with problematic exposures. We also cross-matched the updated \emph{Hipparcos} catalog compiled by \citet{vanLeeuwen2008} to exclude 21 stars with the goodness of fit parameter\footnote{The goodness fit parameter $F2$ is a statistical indicator determining the quality of the measurement. A smaller $F2$ value suggests a robust measurement.} ($|F2|$) $> 5$ \citep{Bhat2022}.
Furthermore, in order to eliminate the contamination from hot subdwarf stars and B-type supergiants,
we excluded stars with $\log{g} > 5.1 \rm cm\ s^{-2}$ and $\leq 3.0 \rm cm\ s^{-2}$\ 
\footnote{Subdwarf-type stars typical $\log{g}$ values of 5.1 to 6.4 $\rm cm\ s^{-2}$ \citep{2009Hebersdb} and typical B-type supergiants with $\log{g}$ $\leq$ 3 $\rm cm\ s^{-2}$ \citep{2007LanztlustyB}.}
and cross-matched the sample to the catalog of hot subdwarf stars identified from LAMOST DR5 to DR7 database by \citet{2019LuoYangping,2020LuoYangping}. 

\begin{figure}
\centering
\includegraphics[scale=0.7]{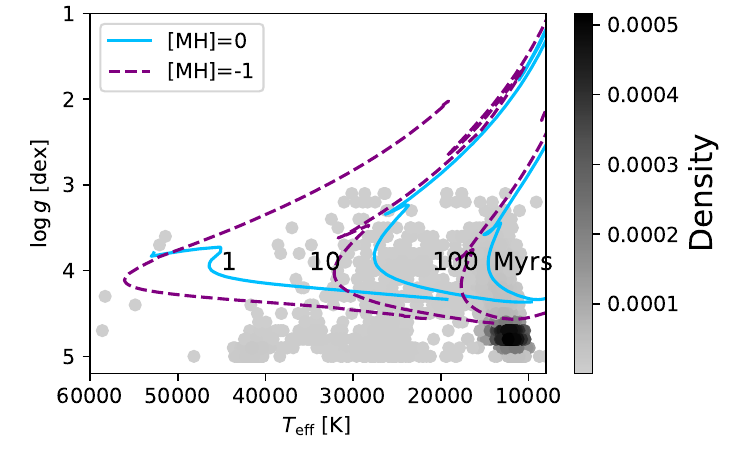}
\caption{The density distribution of our final sample in Hertzsprung–Russell diagram.
The purple and blue lines represent isochrones generated using the Padova and Trieste Stellar Evolutionary Code (PARSEC) for ages of 1, 10, and 100 Myr, with each set corresponding to different metallicity values: [M/H] = −1.0 dex and [M/H] = 0.0 dex, respectively.}\label{fig:HRD}
\end{figure}

We finally reached a sample of 4,432 stars for further study.
Figure~\ref{fig:HRD} shows their locations on the Hertzsprung–Russell diagram, demonstrating that they are dwarfs.
We report the \emph{Gaia} source ID, equatorial coordinate, the collected RV and the associated error from \citet{2021zhangboRV}, the effective temperature ($T_{\rm eff}$), and the projected rotational velocity ($V\sin{i}$) from \citet{2021GYJslam} of each target star in Table~\ref{tab:RV Catalogs}. 
Instead of reporting the individual error for the measured $T_{\rm eff}$, $V\sin{i}$ and $\log{g}$ values of each target star, \citet{2021GYJslam} determined such uncertainties for the whole sample based upon a data-driven algorithm and concluded that the sample stars have error estimation of $\sigma({T_{\rm eff}}) = 2,185$ K, $\sigma(V\sin{i}) = 11$ km s$^{-1}$ and $\sigma(\log{g}) = 0.29$ $\rm cm\ s^{-2}$.

\begin{figure*}
\centering
\includegraphics[scale=0.5]{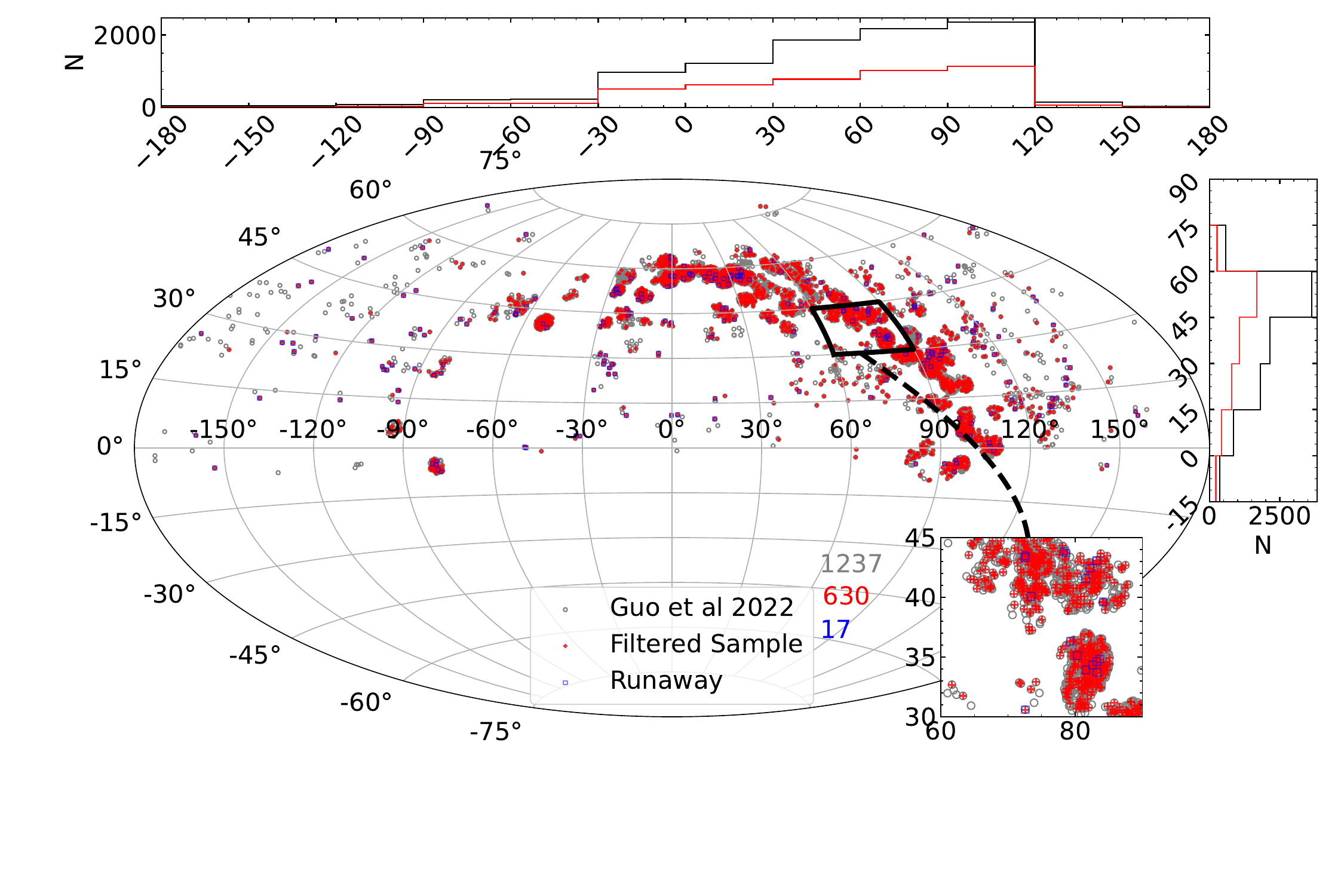}
\caption{The spatial distribution of the initial sample of 9,382 early-type stars released from \citet{2021GYJfb} (gray circles) and the final filtered sample of 4,432 OB stars (red cross). The 229 newly identified runaway candidate stars discussed in Section~\ref{sec:Method} are shown in blue squares. 
The histogram on top of the spatial plot shows the number distribution of stars within each RA interval with a bin size of 30$^\circ$. In this representation, the black steps represent early-type stars released from \citet{2021GYJfb}, and the red steps represent the final filtered sample. A similar number distribution of stars for DEC is shown on the right side of the spatial plot with a bin size of 15$^\circ$. A zoomed-in square subplot on the right lower corner shows the spatial distribution of sample stars distributed with a window size of 30$^\circ \times 15^\circ$ (RA spans from 60$^\circ$ to 90$^\circ$ and 30$^\circ$ to 45$^\circ$ for DEC). Within the selected area, there are 1237 OB stars, 630 filtered stars, and 17 identified runaway candidates. Number counts are labeled adjacent to the subplot. }\label{fig:spatial distribution}
\end{figure*}

\begin{figure}
\centering
\includegraphics[scale=0.6]{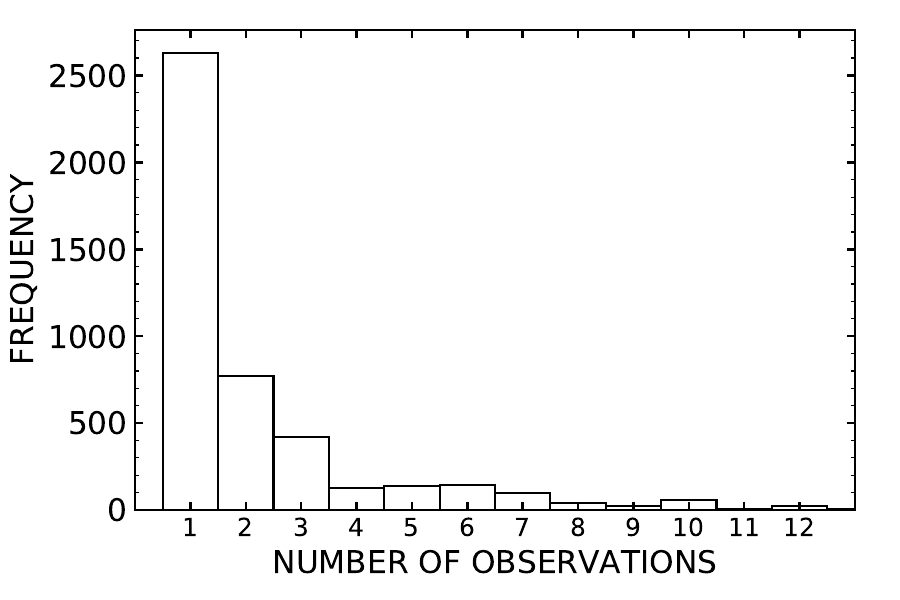}
\caption{The distribution of observed number for the
final filtered sample of 4,432 OB stars (see Sec.~\ref{sec:sample}).}\label{fig:obs_time}
\end{figure}

The spatial location of the 9,382 early-type stars and the final filtered sample of 4,432 stars are shown in Figure~\ref{fig:spatial distribution}. 
For a clear representation of the distribution, the histogram on top of the spatial plot shows the number distribution of stars within each RA interval with a bin size of 30$^\circ$.The black steps represent early-type stars released from \citet{2021GYJfb}, and the red steps represent the final filtered sample. A similar number distribution of stars for DEC is shown on the right side of the spatial plot with a bin size of 15$^\circ$. A zoomed-in square subplot on the right lower corner shows the spatial distribution of sample stars distributed with a window size of 30$^\circ \times 15^\circ$ (RA spans from 60$^\circ$ to 90$^\circ$ and 30$^\circ$ to 45$^\circ$ for DEC). Within the selected area are 1237 OB stars, 630 filtered stars, and 17 identified runaway candidates. Number counts are labeled adjacent to the subplot.
Figure~\ref{fig:obs_time} shows the distribution of the number of observations for the final filtered sample, among which 2,616 stars have a single observation and 1,816 stars have two or more spectra due to the time-domain observing strategy of the MRS survey.

\begin{deluxetable*}{lccccccc}
\tablecaption{The Radial Velocity and atmospheric parameter of the sample stars \label{tab:RV Catalogs}}
\tablewidth{0pt}
\tablehead{
\colhead{\emph{Gaia}} & \colhead{R.A.} & \colhead{Decl.} & \colhead{$V_r$} & \colhead{$\sigma_r$} & \colhead{$T_{\rm eff}$} & \colhead{$V\sin{i}$} & \colhead{$\log{g}$} \\
\colhead{Source ID} & \colhead{(deg)} & \colhead{(deg)} & \colhead{(km s$^{-1}$)} & \colhead{(km s$^{-1}$)} & \colhead{($K$)} & \colhead{(km s$^{-1}$)} & \colhead{($\rm cm\ s^{-2}$)}
}
\startdata
3345118343436889856 & \phn91.7283  & \phs14.1179  & \phs23.4  & 1.8 & 15267 & 35 & 3.6\\
3432295837263529984 & \phn95.4908  & \phs27.0320  & \phs15.5  & 0.9 & 23702 & 64 & 4.3\\
3326717260430731648 & 100.2148 & \phs\phn9.8637 & \phs22.8 & 1.0 & 15724& 36 & 4.5\\
4253180601531525888 & 280.6227 & \phn$-$6.5710  & \phs24.5  & 0.5 & 15594 & 10& 3.8\\
4253380776942510592 & 280.7551 & \phn$-$6.2719  & \phs56.9  & 2.9 & 15682 & 58& 4.0\\
4253188946523867904 & 280.8196 & \phn$-$6.4319  & \phs45.3  & 1.5 & 15158 & 72& 3.5\\
2001666798195652992 & 335.2163 & \phs52.9046  & $-$25.0 & 2.0 & 15111 & 79& 4.0\\
2011627239661167232 & 359.4527 & \phs60.5815  & $-$17.7 & 1.0 & 16415 & 49& 3.7\\
\enddata
\tablecomments{This table is available in its entirety in machine-readable form. The first eight entries are shown here for guidance regarding its format and content.} 
\end{deluxetable*}

\section{Identification of runaway candidate stars} \label{sec:Method}
\begin{figure*}
\centering
\includegraphics[scale=1]{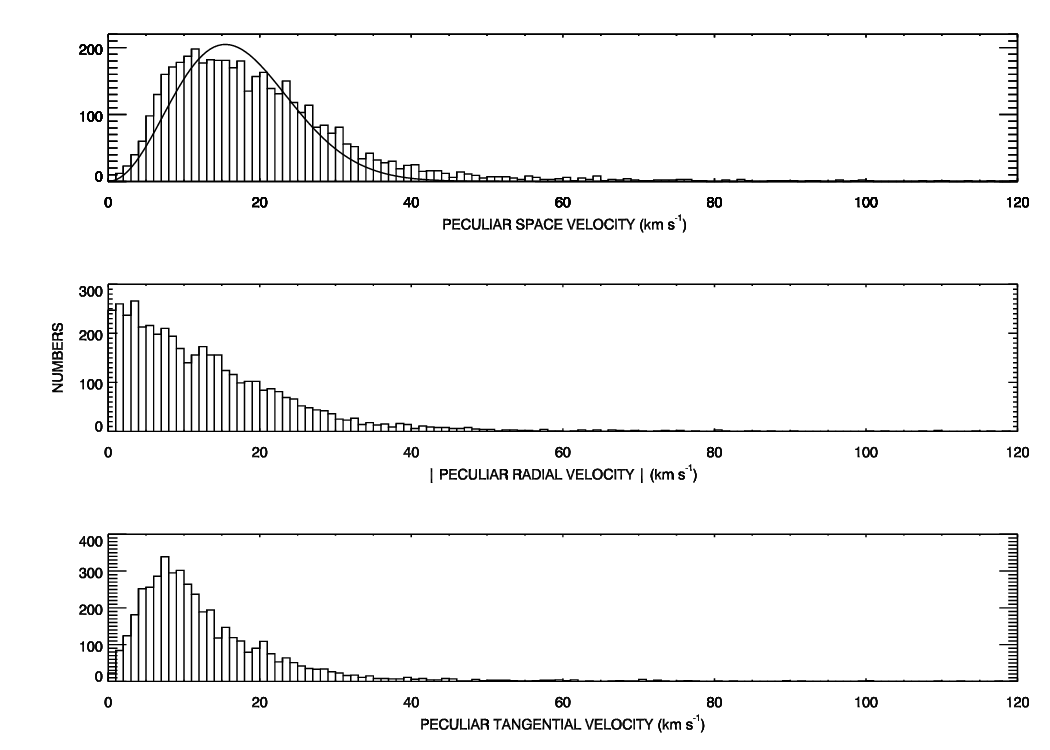}
\caption{From the bottom to the top panel: the number distributions of peculiar tangential velocity, absolute values of peculiar radial velocity, and peculiar space velocity are shown for 4,432 early-type stars identified from the LAMOST MRS DR8.
A Maxwellian velocity distribution fit was applied to the peculiar space velocity and overplotted in the top panel.}\label{fig:Maxwellian distribution}
\end{figure*}

We transformed the equatorial coordinate of the sample stars to the Galactic coordinate system by adopting the north Galactic pole coordinate of ($\alpha_G$, $\delta_G$) = (192.95948$\degr$, 27.12825$\degr$) from the \emph{Hipparcos} Consortium. 
We adopted the solar motion of ($U_\sun$,$V_\sun$,$W_\sun$)=(11.10, 12.24, 7.25) km $\rm s^{-1}$ from \citet{2010Schonrich}, the solar Galactocentric distance $R_0$ = 8.5 kpc, and the circular Galactic rotational velocity $V_c = 220$ km s$^{-1}$ from \citet{1986Kerr}. 
Following the procedures described in \citet{1998Moffat,1999Moffat}, we calculated both the tangential and radial peculiar velocities of the sample stars by subtracting the contribution of the solar motion and the differential Galactic rotation. 
We show the number distributions of calculated peculiar tangential velocity, absolute values of peculiar radial velocity, and peculiar space velocity of the sample stars from the bottom to the top in Figure~\ref{fig:Maxwellian distribution}. 

\cite{1961Blaauw} identified the stars with observed space velocities $>$ 40 km s$^{-1}$ as runaway stars. Since then, a wide exploration of the selection criterion for identifying runaway stars has been carried out by subsequent works based upon kinematical studies from the \emph{Hipparcos} proper motions. 
\citet{Vitrichenko1965} and \citet{Cruz-Gonzalez1974} utilized the peculiar velocity in the radial component of $V_{\rm Rp} \ge 30$ km s$^{-1}$ as the selection criterion to identify Galactic OB runaway stars. \citet{1986Gies} studied a sample of OB runaway stars and found a similar cut-off value for the peculiar radial velocities. \citet{1998Moffat}, \citet{Mdzinarishvili2004}, and \citet{2005Mdzinarishvili} applied a criterion of $V_{\rm Tp} > 42$ km s$^{-1}$ to investigate the runaway population based on the peculiar velocity of the tangential component. In addition to focusing on the selection criterion made from component peculiar motion, \citet{2001Berger} and \citet{deWit2005} have adopted space velocities (with a traditional cut-off value of $V_{\rm Sp} > 40$ km s$^{-1}$), including both radial and tangential components, to search for Galactic runaways. \citet{Tetzlaff2011} noted that applying a lower velocity criterion ($V_{\rm Sp} = \sim30$ km s$^{-1}$) resulted in a 78\% identification efficiency of high-velocity runaway stars, but inevitably included a $\sim9\%$ contamination of low-velocity stars in the identified sample.
In this work, based upon the collected RV measurements from \citet{2021zhangboRV} and astrometric information from {\emph Gaia} DR3, we thus defined a conservative selection criterion by applying  a Maxwellian velocity distribution fit (see the equation below) to the calculated peculiar space velocity of the sample stars, 
\begin{equation}
\label{eq:maxwellian}
f(v) = av^2 \exp(-v^2/b^2)
\end{equation}
where $a = 2.329\pm 0.014$ an d $b = 15.463 \pm 0.032$ km s$^{-1}$. 
We overplotted the Maxwellian velocity fit on the number distribution of the peculiar space velocity in the top panel of Figure~\ref{fig:Maxwellian distribution}. We determined the runaway criterion of $V_{\mathrm Sp}\ > 43$ km s$^{-1}$ from stars with peculiar space velocities greater than the value at $\sim1\%$ of the peak distribution \citep{Berger2001,2022luqianBe}.

Following the determined selection criterion, we identified 229 runaway candidates from the sample collection of 4,432 early-type stars. In Table~\ref{tab:Runaway Catalogs}, we report the equatorial coordinates (R.A.\ and Decl.), the proper motions ($\mu_\alpha$ and $\mu_\beta$), the parallax, and their associated errors collected from the \emph{Gaia} DR3 in columns one to five. Columns six to eight of the table show the calculated peculiar radial velocity ($V_{Rp}$), the peculiar tangential velocity ($V_{Tp}$), the peculiar space velocity ($V_{Sp}$), and their associated errors for the identified candidate stars. We also determined the height of the candidate stars away from the Galactic plane based upon the formula of $z = r\sin{b}+z_\odot$, where $z_\odot = 20.5$ pc is taken from \citet{Humphreys1995}. 
The estimated heights and the associated errors are listed in column 9 of Table~\ref{tab:Runaway Catalogs}.

In our study, we assumed that all sample stars have a galactic rotation velocity (Vc) of 220 km s$^{-1}$. However, some stars are located above the Galactic disk. These objects may have lower galactic rotation velocities than the disk stars, and their peculiar velocities may differ from those obtained in our study \citep{2012Kafle,2019Mroz,2019Eilers,2023Wang}. 
We have not corrected this as the velocity curve at significant z distances is currently unclear. Instead, we have given an assessment of the influence based on the study of \citet{2023Wang}. According to their study, stars with a distance of $|Z| >$ 2.5 kpc have galactic rotational velocities (Vc) that are
20 to 30~km s$^{-1}$ lower than those in the Galactic disk. However, for stars with $|Z|$ = 1.5 kpc and Z = 0 kpc, there are no significant differences between the galactic rotation curves. Therefore, we have only considered stars with $|Z| >$ 1.5 kpc. There are 13 stars total having $|Z| >$ 1.5 kpc, making up 0.29\% of our entire sample. We assumed a linear relationship between Z and Vc (i.e. $Vc=220-f\times|Z|$ where f is a constant and is obtained by setting Vc = 190 km s$^{-1}$ at $|Z| =$ 2.5), and recalculated the galactic rotational velocities of the 13 stars. The results showed that the peculiar velocities of these stars are slightly affected by the adopted value of Vc. However, classifying the same 10 objects out of the 13 stars as runaway stars remains unchanged. We also verified this using Vc=190 km s$^{-1}$ for the 13 objects, which produced similar results.

\begin{figure}
\centering
\includegraphics[scale=0.5]{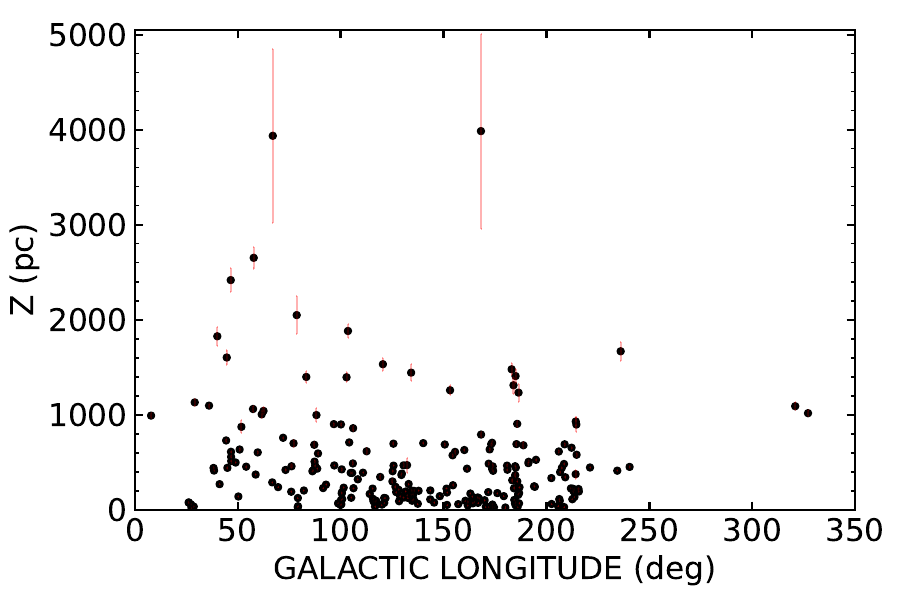}
\caption{The distribution of Galactic heights ( associated errors in red cross) of the 229 newly identified runaway candidate stars as a function of Galactic longitude. Most of the errors are too small to be shown in the figure.}
 \label{fig:z-l-plot}
\end{figure}

In Figure~\ref{fig:z-l-plot}, we show the distribution of Galactic height of the runaway candidate stars as a function of Galactic longitude. 
Our sample has 10 stars with measured Galactic heights greater than 1500 pc. 
However, these stars have large uncertainty in the distance measurements collected from \emph{Gaia} DR3. 
The last three columns in Table~\ref{tab:Runaway Catalogs} listed the Galactic space velocities of the sample stars, in which positive $U$ values indicate the direction moving towards the Galactic center, $V$ points towards the direction of Galactic rotation, and positive $W$ values show the direction moving towards the Galactic North Pole. 
We found a weak correlation between the absolute values of Galactic height $z$ and the $W$ velocity, suggesting stars with large $W$ values would result in higher trajectories away from the Galactic plane. However, a larger sample of runaway stars would be needed to validate such a correlation.

\begin{figure}
\centering
\includegraphics[scale=0.4]{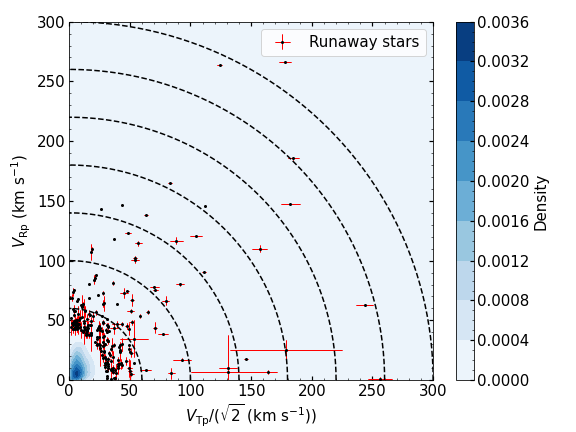}
\caption{The density contour plot shows the distribution of peculiar radial and tangential velocities for the 4,432 OB stars. The velocity entries for the newly identified 229 runaway candidates are overplotted in the figure in black dot and their associated errors in red cross. 
Error bars (in red) displaying the uncertainties of peculiar velocities are included for the 229 identified runaway candidates.
The dashed arcs represent the constant values of $(V_{Rp})^{2}+(V_{Tp}^{2})/2$ ranging from 60 to 260 km s$^{-1}$, with an increment of 40 km s$^{-1}$. }\label{fig:Peculiar RV distribution}
\end{figure}

\begin{figure}
\centering
\includegraphics[scale=0.6]{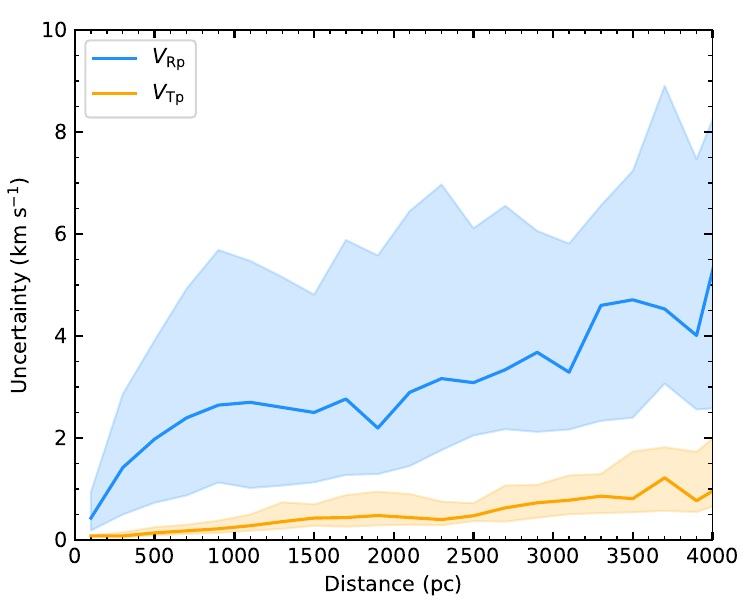}
\caption{Uncertainty distributions of the peculiar radial (blue) and tangential (orange) velocity measurements are shown as a function of distance. The median values are plotted in solid lines, and the 16th to 84th percentiles of the error distributions are shown in the shaded regions. Since 98$\%$ of our sample stars have distances within 4 kpc, the distributions are plotted for such range only.}\label{fig:Uncertainty spider}
\end{figure}

Figure~\ref{fig:Peculiar RV distribution} shows the density contour plot of the peculiar radial and tangential velocity distribution for a total of 4,432 OB stars. The velocities of identified runaway candidates (black dot) and their associated errors (red cross) are overplotted in the figure.
The distribution of peculiar velocities show an asymmetric feature in Figure~\ref{fig:Peculiar RV distribution}. The peculiar velocity measurements depend on individual target stars' proper motion and distance. We reviewed available past runaway works and noticed that such asymmetric feature also appears in the ones from \citet{Tetzlaff2011,1998Moffat,Berger2001}. \citet{Gaia2023} utilize about 580,000 OB stars to investigate the Milky Way disc features and note that for stars within a distance of $\sim$4 kpc, the uncertainty of velocity in the radial direction is more scattered compared to the ones from the tangential component (see the top panel of their Fig.~10). In our sample, about 98\% of the target stars shown in Figure~\ref{fig:Peculiar RV distribution} have given distance within $\sim4$ kpc. We show the uncertainty distribution in the peculiar radial velocity and tangential velocity measurements as a function of distance in Figure~\ref{fig:Uncertainty spider}. The errors in the radial direction display a larger scatter than those of the tangential component.
Such an asymmetry feature may suggest that Galactic velocity diffusion has a more noticeable impact on velocity measurements in the radial direction.

\subsection{The impact of limited RV measurements}

As introduced in Section 1, a significant fraction of early-type stars are binaries and their orbital motions may result in RV of hundreds of km/s. On the other hand, more than 3/4 of our sample has only one or two radial velocity measurements as shown in Figure 5. As a consequence, the measured values of RV may be caused by orbital motion rather than space motion of the binary. Even though we use *space velocity* here to perform our study, we can not correct the contribution from binary orbital motion due to the limited number of  RV observations. 
However, we may imagine that it is more likely that the massive component of a binary has been observed (which is close to the mass center of the binary) and the contribution of orbit motion to RV (and space velocity) is not very large, depending on the mass ratio.

\begin{figure}
\centering
\includegraphics[scale=0.6]{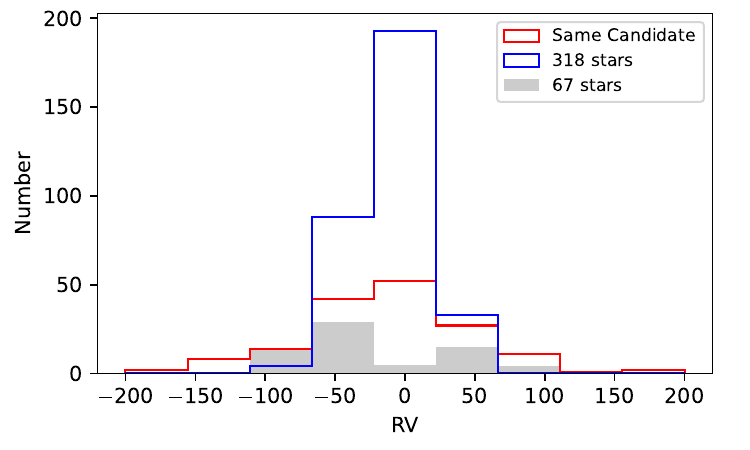}
\caption{The RV distribution of two catalogs.
The 162 stars with Flag=S in both Table~\ref{tab:Runaway Catalogs} and Table~\ref{tab:Runaway Catalogs Rt} are indicated by the red step. The 67 stars are only present in Table~\ref{tab:Runaway Catalogs} with Flag=N, marked by the gray shade, while the 318 stars are exclusively shown in Table~\ref{tab:Runaway Catalogs Rt} with Flag=N, represented by the blue step.
}
 \label{fig:Vtp_run.pdf}
\end{figure}

We made a test using the transverse velocity instead of the space velocity as the indicator of runaway stars and identified 480 runaway candidates, nearly double the number obtained from the space velocities (229 stars). 
There are 162 stars found using both methods, with an overlap rate of 71\%(162/229).
We present the RV distribution of the 162 objects and compare it with the left 67 (=229-162, with Flag=N in Table~\ref{tab:Runaway Catalogs}) and 318 (=480-162, with Flag=N in Table~\ref{tab:Runaway Catalogs Rt}) samples in two catalogs, as shown in Fig.~\ref{fig:Vtp_run.pdf}. The common sample comprising 162 stars (in red) has a wide range of RV values, while the 67 objects only chosen by their space velocity (in black) have moderate RVs. 
It is difficult to estimate the contribution of binary motion to the RVs for the 67 stars presently. 
We checked their spectra and found that 6 stars exhibit a weak H$_{\alpha}$ asymmetric absorption core feature and one has double lines in all absorption lines. 
We have annotated the 7 objects in Table~\ref{tab:Runaway Catalogs} with Flag=B. 
On the other hand, for the 318 objects chosen solely by their transverse velocities, 
the distribution of RV peaks at around 0 and shows good symmetry as expected.
Even though the unidentified binaries may inflate the peculiar radial velocities, we can only address this issue in future work by excluding such binaries through the incorporation of more observational data.

The comparison above indicates that the 3D space velocity criterion is more rigorous and effective in identifying runaway stars than the transverse velocity criterion. However, due to the limited number of RV measurements in our sample, a small fraction of binaries may have been mixed in. As a comprehensive reference, we also present the runaway candidates identified through the transverse velocity criterion in Table~\ref{tab:Runaway Catalogs Rt}. The common stars identified by the two methods are marked with Flag=S in Table~\ref{tab:Runaway Catalogs} and Table~\ref{tab:Runaway Catalogs Rt}.

\begin{deluxetable*}{ccccccccccccc}
\rotate
\tablecaption{The Peculiar Velocities of the 229 Runaway Candidate Stars Identified from the LAMOST MRS DR8 Sample \label{tab:Runaway Catalogs}}
\tabletypesize{\scriptsize}
\tablewidth{0pt}
\tablehead{
 \colhead{R.A.}   & \colhead{Decl.}  & 
 \colhead{$\mu_\alpha$} & \colhead{$\mu_\beta$} & \colhead{Parallax} & \colhead{$V_{Rp}$} & \colhead{$V_{Tp}$}  & \colhead{$V_{Sp}$} & \colhead{$z$} & \colhead{$U$} & \colhead{$V$} & \colhead{$W$} & \colhead{Flag} \\
 \colhead{(deg)}   & \colhead{(deg)}  & 
 \colhead{(mas yr$^{-1}$)} & \colhead{(mas yr$^{-1}$)} & \colhead{(mas)} & \colhead{(km s$^{-1}$)} & \colhead{(km s$^{-1}$)}  & \colhead{(km s$^{-1}$)} & \colhead{(pc)}& \colhead{(km s$^{-1}$)} & \colhead{(km s$^{-1}$)}  & \colhead{(km s$^{-1}$)} & \colhead{}
 }
 \startdata
\phn\phn6.0719    & 56.7296 & $-$2.5310    $\pm$ 0.0150  &  \phn$-$2.7660  $\pm$ 0.0180  &  0.3167 $\pm$ 0.0177   &  \phs53.47 $\pm$ \phn3.85   &  \phn27.87   $\pm$ \phn1.60    &  \phn60.30    $\pm$ \phn4.17  & \phn347.5 $\pm$ \phn18.3 & \phs42.85 $\pm$ \phn2.07 & \phs261.62 $\pm$ 1.92  & $-$31.17 $\pm$ 2.32 & S\\
\phn19.8459   & 58.3056 & $-$1.6910    $\pm$ 0.0100  &  \phn$-$0.7440  $\pm$ 0.0110  &  0.3402 $\pm$ 0.0133   &  \phs69.15 $\pm$ \phn2.67   &  \phn\phn3.58    $\pm$ \phn0.25   &  \phn69.24   $\pm$ \phn2.68  & \phn244.3 $\pm$ \phn\phn8.8 & \phs10.10 $\pm$ \phn1.14 & \phs270.98 $\pm$ 1.29  &  \phn$-$8.07 $\pm$ 0.45 & N \\
114.3282  & 45.1335 & \phs1.2620   $\pm$ 0.0170  &  $-$49.3980 $\pm$ 0.0150  &  0.6525 $\pm$ 0.0182   &  $-$62.65  $\pm$ \phn1.02   &  344.46  $\pm$ \phn9.61   &  350.11  $\pm$ \phn9.66  & \phn707.1 $\pm$ \phn19.2 & \phn$-$5.77 $\pm$ \phn2.26 & $-$120.22 $\pm$ 9.60 & $-$82.20 $\pm$ 2.03 & S\\
273.8046  & 39.7094 & $-$0.0980    $\pm$ 0.0270  &  \phn$-$0.9840  $\pm$ 0.0310  &  0.1019 $\pm$ 0.0238   &  \phn\phs6.56  $\pm$ 30.99  &  184.97  $\pm$ 43.24  &  185.09  $\pm$ 53.20 & 3936.9 $\pm$ 914.7 & \phs42.20 $\pm$ 10.13 &  \phs189.90 $\pm$ 4.91 & $-$13.88 $\pm$ 3.66 & S\\
359.8080  & 63.8195 & $-$3.8280    $\pm$ 0.0150  &  \phn$-$0.7450  $\pm$ 0.0140  &  0.3496 $\pm$ 0.0155   &  \phs37.88 $\pm$ \phn2.34   &  \phn23.44   $\pm$ \phn1.08   &  \phn44.55   $\pm$ \phn2.58  & \phn\phn96.8  $\pm$ \phn\phn3.4 & \phs57.96 $\pm$ \phn2.03 & \phs256.74 $\pm$ 1.08 & \phs\phn7.80 $\pm$ 0.67 & N\\
\enddata
\tablecomments{This table is available in its entirety in machine-readable form. The first five entries are shown here for guidance regarding its format and content.}
\end{deluxetable*}

\begin{deluxetable*}{ccccccccccccc}
\rotate
\tablecaption{The 480 Runaway Candidate Stars based on transverse velocity \label{tab:Runaway Catalogs Rt}}
\tabletypesize{\scriptsize}
\tablewidth{0pt}
\tablehead{
 \colhead{R.A.}   & \colhead{Decl.}  & 
 \colhead{$\mu_\alpha$} & \colhead{$\mu_\beta$} & \colhead{Parallax} & \colhead{$V_{Rp}$} & \colhead{$V_{Tp}$}  & \colhead{$V_{Sp}$} & \colhead{$z$} & \colhead{$U$} & \colhead{$V$} & \colhead{$W$} & \colhead{Flag}\\
 \colhead{(deg)}   & \colhead{(deg)}  & 
 \colhead{(mas yr$^{-1}$)} & \colhead{(mas yr$^{-1}$)} & \colhead{(mas)} & \colhead{(km s$^{-1}$)} & \colhead{(km s$^{-1}$)}  & \colhead{(km s$^{-1}$)} & \colhead{(pc)}& \colhead{(km s$^{-1}$)} & \colhead{(km s$^{-1}$)}  & \colhead{(km s$^{-1}$)} & \colhead{}
 }
 \startdata
\phn\phn6.0719    & 56.7296 & $-$2.5310    $\pm$ 0.0150  &  \phn$-$2.7660  $\pm$ 0.0180  &  0.3167 $\pm$ 0.0177   &  \phs53.47 $\pm$ \phn3.85   &  \phn27.87   $\pm$ \phn1.60    &  \phn60.30    $\pm$ \phn4.17  & \phn347.5 $\pm$ \phn18.3 & \phs42.85 $\pm$ \phn2.07 & \phs261.62 $\pm$ 1.92  & $-$31.17 $\pm$ 2.32 & S\\
108.8516  & 0.789   & $-$3.7290  $\pm$ 0.0350  &  1.9970      $\pm$ 0.0300  &  0.4931 $\pm$ 0.0317   &  14.87 $\pm$ 2.56         &  26.87   $\pm$ 1.78   &  30.71   $\pm$ 3.12  &  221.63 $\pm$ 1.2    &   31.17$\pm$ 0.41&   232.66$\pm$ 0.79&   9.36  $\pm$ 0.11& N\\
114.3282  & 45.1335 & \phs1.2620   $\pm$ 0.0170  &  $-$49.3980 $\pm$ 0.0150  &  0.6525 $\pm$ 0.0182   &  $-$62.65  $\pm$ \phn1.02   &  344.46  $\pm$ \phn9.61   &  350.11  $\pm$ \phn9.66  & \phn707.1 $\pm$ \phn19.2 & \phn$-$5.77 $\pm$ \phn2.26 & $-$120.22 $\pm$ 9.60 & $-$82.20 $\pm$ 2.03 & S\\
273.8046  & 39.7094 & $-$0.0980    $\pm$ 0.0270  &  \phn$-$0.9840  $\pm$ 0.0310  &  0.1019 $\pm$ 0.0238   &  \phn\phs6.56  $\pm$ 30.99  &  184.97  $\pm$ 43.24  &  185.09  $\pm$ 53.20 & 3936.9 $\pm$ 914.7 & \phs42.20 $\pm$ 10.13 &  \phs189.90 $\pm$ 4.91 & $-$13.88 $\pm$ 3.66& S\\
359.0649  & 62.73   & $-$3.8470    $\pm$ 0.0120  &  $-$1.1270      $\pm$ 0.0140  &  0.3017 $\pm$ 0.0122   &  $-$9.71 $\pm$ 3.29         &  26.51   $\pm$ 1.11   &  28.23   $\pm$ 3.47  &  51.3 $\pm$ 1.2    &    30.08$\pm$ 2.59&   200.60$\pm$ 2.44&   -7.28  $\pm$ 1.28& N\\
\enddata
\tablecomments{This table is available in its entirety in machine-readable form. The first five entries are shown here for guidance regarding its format and content.}
\end{deluxetable*}

\section{Properties of the runaway stars} \label{sec:Result and D} 

We identified 229 runaway candidate stars from a sample of 4,432 early-type stars from LAMOST MRS DR8. 
Based upon the previously measured atmospherical parameters of these runaway candidate stars from \citet{2021GYJslam}, we discuss their statistical properties below. 

\subsection{Completeness and Bias}
The early-type stars drawn from the LAMOST MRS survey have followed the footprints illustrated in Figure~2 of \citet{2020LiuChao}, and the survey selects only a small fraction of targets to observe. Here, we quantitatively discuss the completeness of the runaway candidate population identified from the MRS database through visual inspection. We arbitrarily selected a rectangular region (R.A.\ spans from 90$^\circ$ to 120$^\circ$ and Decl.\ from 45$^\circ$ to 60$^\circ$) in Figure~3 of Section~\ref{sec:sample} from the LAMOST MRS database. Such an area overlaps with the spatial distribution of our initial working sample and contains a total collection of 16,130 stars. We visually inspected each spectrum to identify the OB stars from the selected sample. Through further filtering (reject entries with $S/N < 50$ and {\tt ruwe} $> 1.4$), we arrived at a collection of 46 OB stars. Among these visually identified OB stars, 34 appear in the final sample of 4,432 early-type stars identified from the $EW$ indices in Section~\ref{sec:final sample} but missed 12 detections, and thus corresponds to a completeness of 74$\%$ (34/46), consistent with the results of \citet{2019LiuZhicun}. We inspected the spectra of the 12 missed OB detections and measured their associated $EW$ indices. The measurements were situated outside the selection criteria of OB stars from \citet{Guo2022} (see their Fig.~1), suggesting that the adopted $EW$ approach to identify OB from \citet{Guo2022} is more stringent and conservative.

Five runaway candidate stars are within the selected rectangular area, as identified in Section~\ref{sec:Method}. To check whether any runaway stars were missing from the visually identified OB samples, we applied the procedures from Section~\ref{sec:Method} to search for runaway candidate stars in the missing sample of 12 OB stars, yielding two new detections. Compared to the original detection of five runaway candidates from the 34 OB stars, we reach a completeness of $\sim71\%$ (5/7). As described in Section~\ref{sec:Method}, identifying runaway candidate stars depends on proper motion and RV measurements of target stars. We thus applied a series of Kolmogorov$-$Smirnov (K$-$S) tests to the distribution of peculiar velocities (including $V_{Rp}$, $V_{Tp}$, and $V_{Sp}$) and RV measurements for the OB samples identified from visual inspection and the ones from the $EW$ indices to evaluate the impact of the selection bias on the runaway sample stars. The two samples have $p$ values near unity, indicating that they are drawn from the same underlying distribution.  
This result suggests that the peculiar velocity and RV distributions of the two samples are not affected by how we selected the OB samples and, thus, the runaway population. 
While it is inevitable that we miss some OB detections in the sample when adopting the EW indices method, our methodology itself did not introduce bias into the selection of the final runaway sample in choosing OB star samples.

Furthermore, we divided the sample into two/three groups according to their distances. The groups had about an equal number of samples but were located at varying distance ranges. We then use the method in Section~\ref{sec:Method} to define the criterion of the runaway stars. The results show that the criterion for dividing the sample into two groups are 41.3 km s$^{-1}$ and 44.4 km s$^{-1}$, and for dividing it into three groups are 40.3 km s$^{-1}$, 43.3 km s$^{-1}$ and 44.4 km s$^{-1}$, respectively. This indicates that our sample's incompleteness does not introduce bias into the selection of the runaway sample.

We caution that the completeness and selection bias discussed here refers to the number of runaway candidate stars identified from the LAMOST MRS database, and it is not the intrinsic completeness for the OB runaway population. 

\subsection{The $T_{\rm eff}$ distribution of identified runaway population }
\begin{figure}
\centering
\includegraphics[scale=0.6]{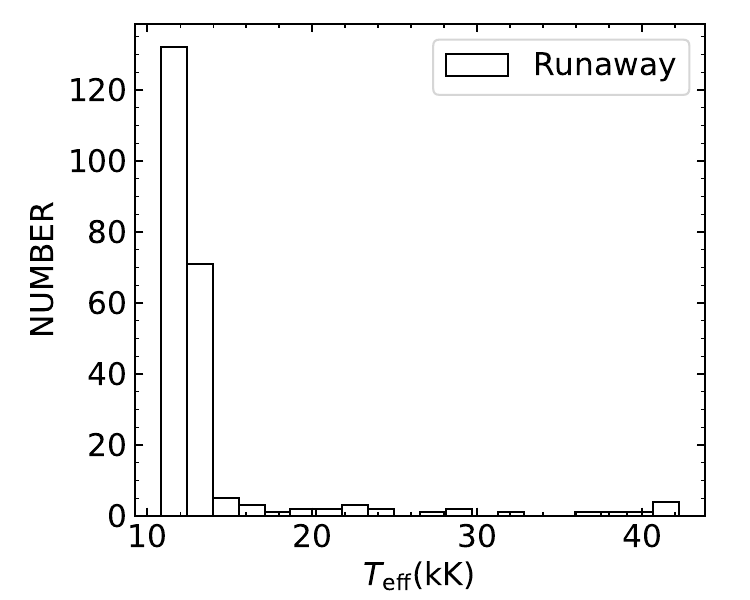}
\caption{The number distribution of measured effective temperature ($T_{\rm eff}$) of the identified 229 runaway candidate stars.}\label{fig: hist T}
\end{figure}

Figure \ref{fig: hist T} displays the number distribution of $T_{\rm eff}$ values for the identified runaway population.
Based upon the $T_{\rm eff}$ values obtained from the stellar parameter survey by \citet{2021GYJslam}, we notice that in our runaway sample, $\sim3\%$ (8 out of 229 stars) of them are hot stars with estimated $T_{\rm eff} > 30,000$ K (B0), $\sim2\%$ (5 stars) of the population are relatively cooler stars with $T_{\rm eff} < 11,400$ K (B8). Most runaway stars ($95\%$) in the new sample have determined $T_{\rm eff}$ between 11,400 K and 30,000 K. 

\subsection{The $V\sin{i}$ distribution of identified runaway population} 
\begin{figure}
\centering
\includegraphics[scale=0.6]{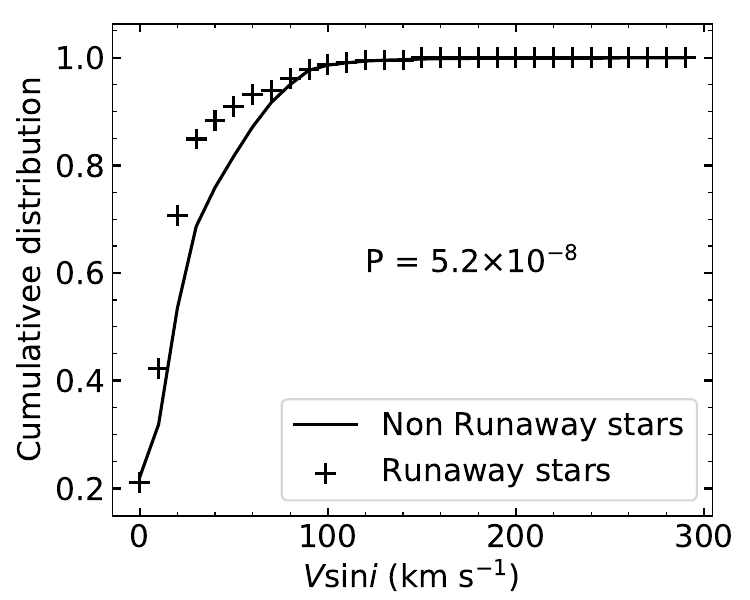}
\caption{The cumulative distribution functions of $V\sin{i}$ values for both runaway candidates (cross sign) and non-runaway population (solid curve) for the OB sample stars from LAMOST MRS. 
The P value is obtained from the KS test.}\label{fig:ks_vsini}
\end{figure}

\begin{figure}
\centering
\includegraphics[scale=0.6]{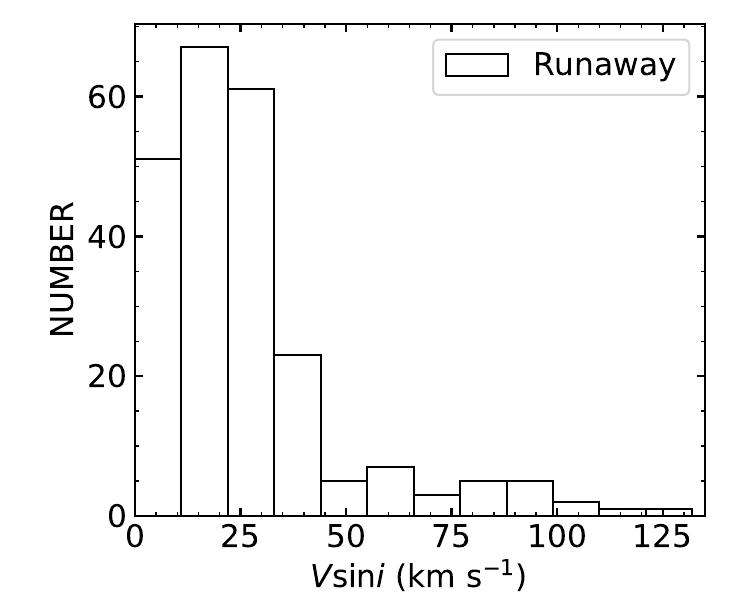}
\caption{The number distribution of the projected rotational velocity ($V\sin{i}$) of the identified 229 runaway candidates.}\label{fig:hist vsini}
\end{figure}

\begin{figure}
\centering
\includegraphics[scale=0.6]{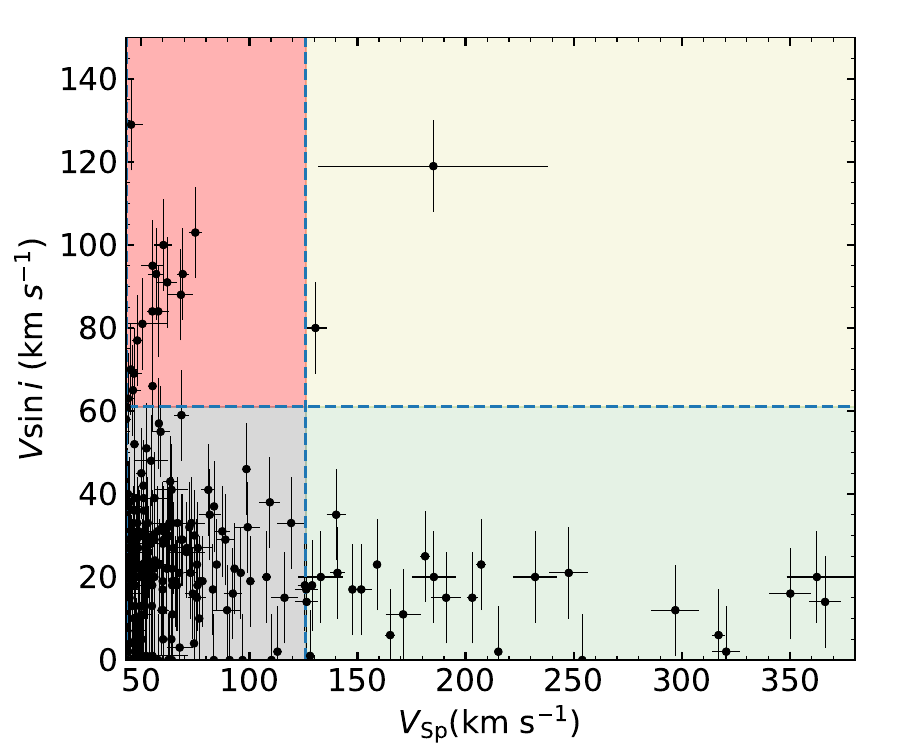}
\caption{Distribution of $V\sin{i}$ and peculiar space velocity ($V_{\mathrm Sp}$) of the newly identified runaway candidates.}\label{fig:Vsini-Vsp}
\end{figure}

We are also interested in studying the projected rotational velocity ($V\sin{i}$) distribution of the identified runaway stars.
We first applied a cumulative distribution function (CDF) calculation to the $V\sin{i}$ values of the identified runaway population (229 stars, labeled as cross sign) and non-runaway stars (4,203 stars, solid curve) in the working sample, and such results are shown in Figure~\ref{fig:ks_vsini}. 
The small P-value\footnote{A smaller P-value indicates that the compared samples differ significantly and may not draw from the same distribution.} ($<1\%$) obtained from the Kolmogorov-Smirnov test suggests that these two populations may have different parent distributions.    

Figure~\ref{fig:hist vsini} displays the number distribution of $V\sin{i}$ values for the identified runaway population. In order to characterize the $V\sin{i}$ values of the runaway candidate stars, we then applied a Maxwellian fitting to the measurements by defining rapid rotators as stars with $V\sin{i}$ values higher than 3$\sigma$ level (where $\sigma=22$ km s$^{-1}$ is the most probable velocity of the $V\sin{i}$ distribution), i.e., this corresponds to $V\sin{i}$ at 1\% of the peak distribution. We obtained the fitting parameters of $a=0.494 \pm 0.028$ and $b=21.861 \pm 0.417$ km s$^{-1}$. 
In our sample, most runaway stars are slow rotators (210 out of 229, corresponds to 92\%) with $V\sin{i}$ smaller than the defined value.

The distribution of the measured peculiar space velocities of the runaway stars is asymmetric. We also applied a Maxwellian fit to the measurements and obtained a value of $V_{\rm Sp} = 126$ km s$^{-1}$ at 1\% of its peak distribution, and the associated fitting values of $a=0.218 \pm 0.004$ and $b=43.818 \pm 0.249$ km s$^{-1}$. Figure~\ref{fig:Vsini-Vsp} presents the $V\sin{i}$ distribution of the runaway candidates to their peculiar space velocity ($V_{\mathrm Sp}$). Based upon the fitted Maxwellian distributions (shown as the blue dashed line on Figure~\ref{fig:Vsini-Vsp}), we arbitrarily group the runaway sample stars into four categories, including slow rotators with small $V_{\rm Sp}$ values (183 stars, gray region), most of the sample stars reside in this region; rapid rotators with small $V_{\rm Sp}$ values (17 stars, red region); slow rotators with large $V_{\rm Sp}$ values (27 stars, light green region); and rapid rotators with large $V_{\rm Sp}$ values (yellow-shaded area). There are only two runaway candidate stars marginally located within the yellow region, which is known as a ``runaway desert". This phenomenon is also shown in \citet{2022Sana} (see their Fig.~4).
\citet{2022Sana} investigated a sample of O-type runaway stars in 30 Doradus and found that rapidly rotating stars dominate the sample.
The authors argue that $V\sin{i}$ values may express hints associated with the formation scenarios of the runaway population, i.e.\ fast moving (large $V_{\rm{Sp}}$ values), slow spinning (low $V\sin{i}$ values) stars are likely formed through the DES scenario, while slow-moving (small $V_{\rm{Sp}}$ values) but rapid rotators (high $V\sin{i}$ values) might be from the BSS.

In the case of BSS scenario, rapidly rotating runaways may originate from supernovae explosions as a result of past close binary interactions \citep{Hoogerwerf2001}. In such a case, runaway stars may be expelled either by a Type Ia SN or a core-collapse SN. The theoretical study by \citet{Han2008} suggested that the companion stars in binary progenitor systems of Type Ia supernovae generally have a mass below $3-5$\,M$_{\odot}$ at the moment of SN explosion \citep[see also Figure~7 of][]{Liu2018}, such mass range is below the known mass for O-type stars ($>17\,M_\odot$). This implies that O-type runaway stars are not likely formed from the Type Ia SN channel. 
Alternatively, the stars with an initial mass of $>8\,M_{\odot}$ but less than $40\,M_{\odot}$ are usually thought to experience core-collapse supernovae \citep[e.g.][]{2003HegerSNII,2008PoelarendsSNII}.

We notice that a big fraction of our sample stars (91\%) are slow rotators with measured $V\sin{i}$ values below the defined criterion of 3$\sigma$ velocity level. Recent theoretical simulations \cite[e.g.][]{2013pan,2022zengyaotian,2023LiuSNII} suggest that when a supernova explodes in a binary system, the interaction between supernova ejecta and its companion star would expect to strip off some material from the companion surface. 
The stripped material takes away some angular momentum of the star, and the star puffs up due to the shock heating during the ejecta-companion interaction. As a result, the rotation of the companion star could drop after the SN explosion. 
This might help to slow down the rotation of the runaway stars if they are produced from the BSS.

Very recently, \citet{2023gaianewrunaway} presented a catalog of O- and Be-type runaway stars based on Gaia DR3. They suggest that the incidence rate of runaway stars is higher for O-type stars than the Be population, indicating that the dynamic ejection scenario (DES) is more favored than the binary supernova scenario (BSS).

\subsection{The U-V-velocity diagram of runaway stars}
\begin{figure}
\centering
\includegraphics[scale=0.6]{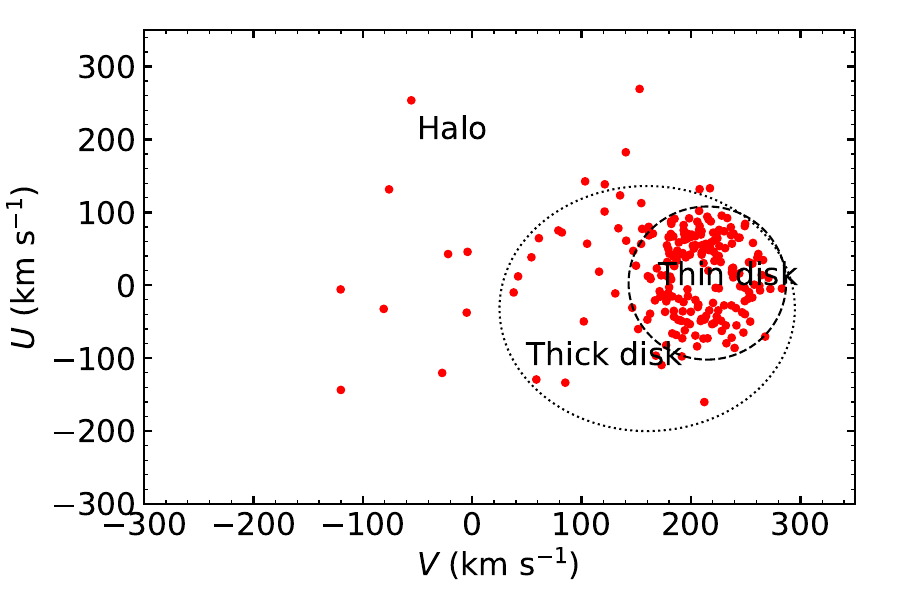}
\caption{$U-V$ velocity diagram of the 229 runaway candidate stars identified from the LAMOST MRS DR8.}  
 \label{fig:UV plot}
\end{figure}

The $U-V$-velocity diagram is a well-established classical method to investigate the kinematic features of the stellar population and to identify their population membership (i.e.\ Galatic thin or thick disk stars, and Galactic halo stars) based on their distribution on the $U-V$-diagram \citep{2006Pauli}. 
Based upon the kinematic features of the runaway stars, with careful modeling of the Galactic potential, the trajectories of the runaway stars can be calculated\citep{Bovy2015,Irrgang2021}. Then, through Monte Carlo simulations or minimization fitting procedures (e.g., \ \citealt{Bhat2022}), the original sites of the runaway population can be traced back by minimizing the distances between computed trajectories.
We show the distribution of the $U$ and $V$ velocities of the runaway population in Figure~\ref{fig:UV plot}.
The dashed and dotted contours represent the $U-V$ velocity distribution for the thin-disk and thick-disk population from \citet{2006Pauli}, and stars residing outside the dotted contours are classified as Galactic halo stars. 
The distribution suggests that 81\% (186 out of 229) of the runaway stars share similar kinematics with the thin disk, while about 19\% of runaway stars exhibit kinematics similar to those of the thick disk (28 stars) or the Galactic halo (15 stars).
A firmed confirmation of the population membership will need further investigations on individual stars' orbital parameters (e.g., the $z-$component of angular momentum or orbital eccentricity) \citep{Pauli2003,2006Pauli}.

\subsection{Binary candidate star}
\begin{figure*}[h]
\gridline{
        \fig{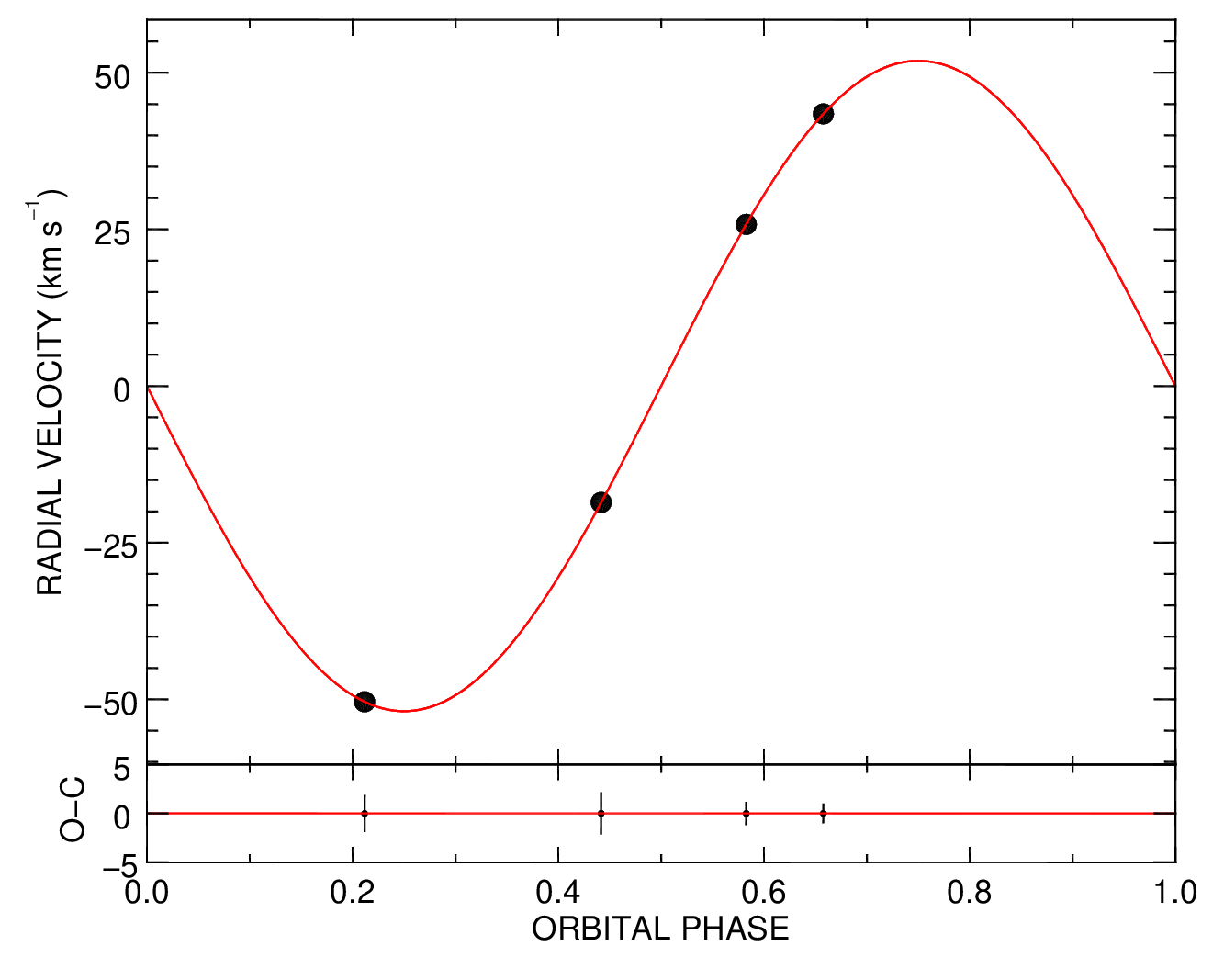}{0.5\textwidth}{(a) \emph{Gaia} Source ID~454293469890174976 }
        \fig{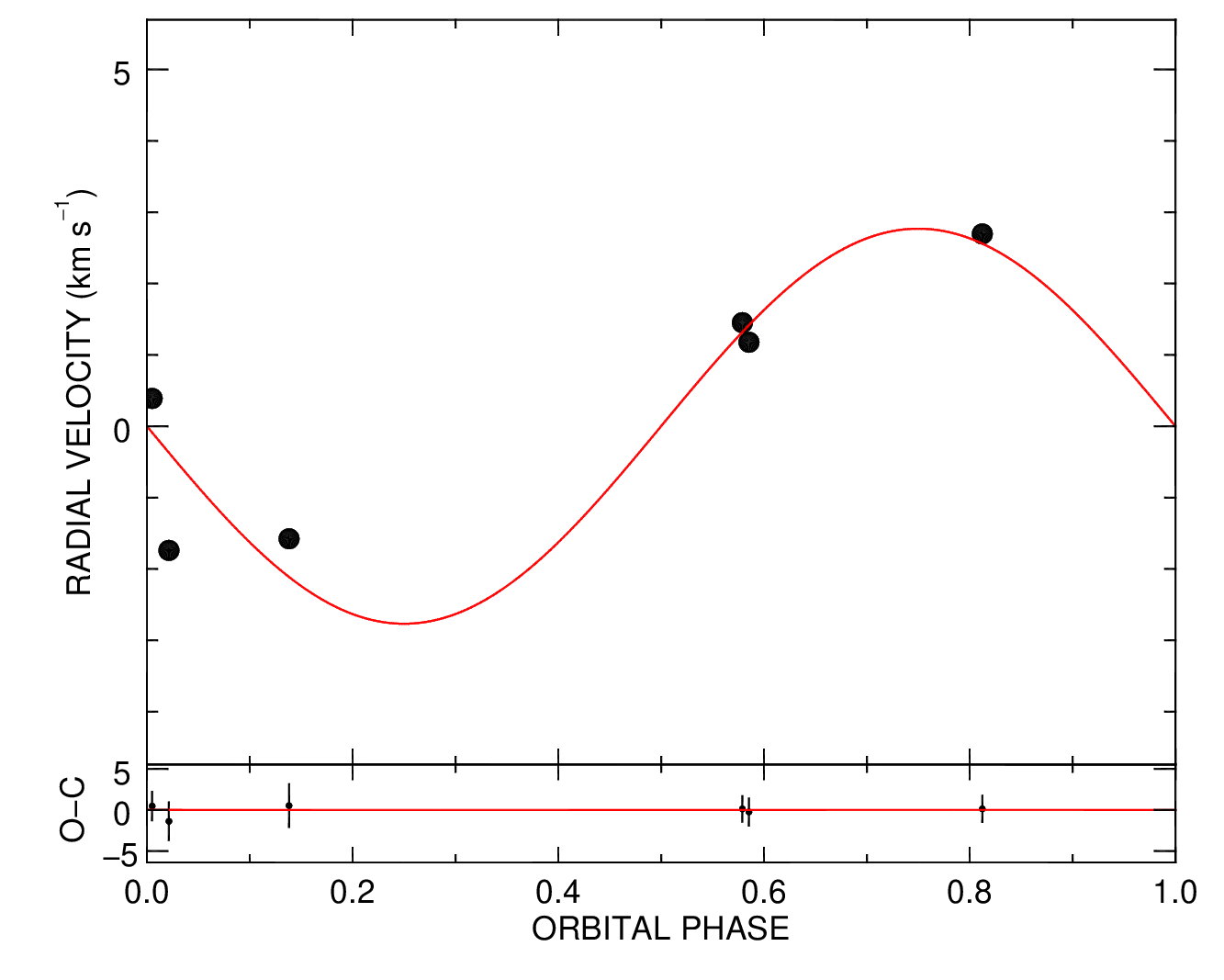}{0.5\textwidth}{(b) \emph{Gaia} Source ID~3453566198941548160 }
        }
\caption{(a): Top panel shows the RV curve for the binary candidate systems \emph{Gaia}~Source~ID~454293469890174976. RV measurements were collected from prior studies by \citet{2021zhangboRV}. Bottom panel displays observed RVs minus the calculated values obtained from the circular orbital fitting using a global minimization technique from \citet{Iglesias-Marzoa2015}. RV curve for \emph{Gaia} Source ID~3453566198941548160 (b) is shown in the same format as (a). }
\label{fig:RV-binary}
\end{figure*}

\placefigure{fig:RV-binary}

Binary interaction plays an important role in tracing the formation scenario of runaway stars \citep{1961Blaauw}. We thus are also interested in exploring the binarity property of our newly identified runaway sample. Based upon the collected RV measurements of the sample stars, we noticed that two of the targets, \emph{Gaia}~Source~ID~454293469890174976 ($V_{\rm Sp} = 46.78 \pm 6.14$ km s$^{-1}$) and \emph{Gaia}~Source~ID~3453566198941548160 ($V_{\rm Sp} = 50.59 \pm 11.24$ km s$^{-1}$) display significant velocity variation (Fig.~\ref{fig:RV-binary}). 
We adopted a global minimization convergence technique developed by \citet{Iglesias-Marzoa2015} to explore the orbital fits to the RVs by assigning equal weight to each RV measurement. We found that the best tentative solution was achieved from a circular orbit fit. We present the fitted RV curve in red in Figure~\ref{fig:RV-binary} and report the estimated single-lined orbital solution of these systems in Table~\ref{tab:orbit_binary}. 
As noted by \citet{Hoogerwerf2001}, binary runaway stars may be formed through the BSS scenario, in which the final binary system consists of a runaway star being expelled by the SN explosion and a compact remnant star, such as a neutron star or a black hole. The observed OB stars with a neutron star companion, e.g., \ High Mass X-ray Binaries (HMXBs), have typical orbital periods ranging from 10 to 300 days \citep{Reig2011}. Our two binary candidate systems have orbital periods between 40 to 60 days, which are well situated in the known range. However, we caution that our preliminary orbital fits are based on the limited number of RV observations. The secondary companion might not be a compact star. Additional RV measurements will be needed to confirm the binary properties of these systems, especially the nature of the companion star. 

\begin{deluxetable}{lrr}[h!]
\tablecaption{Circular orbital elements of the two binary candidate systems \label{tab:orbit_binary}}
\tablewidth{0pt}
\tablehead{
\colhead{Element} & \multicolumn{2}{c}{Value} \\
    & \multicolumn{2}{c}{} \\
\cline{2-3} & A & B
}
\startdata
$P$ (days)		& 40.17 $\pm$ 0.016 &  61.13 $\pm$ 0.75\\
$T_\mathrm{sc}^a$ (HJD$-$2,400,000)	& 58431.69 $\pm$ 0.26  & 58445.50 $\pm$ 3.26\\
$K_1$ (km s$^{-1}$)		& 51.9 $\pm$ 0.9  & 2.8 $\pm$ 0.8\\
$\gamma$ (km s$^{-1}$)	& 20.9 $\pm$ 0.7  & $-$29.3 $\pm$ 0.8\\
$f(m)$ ($M_\odot$) & 0.581 $\pm$ 0.7   & 0.00013 $\pm$ 0.00011\\
$a_1\sin i$ ($R_\odot$)	&  41.2 $\pm$ 0.7  & 3.3 $\pm$ 0.9\\
rms (km s$^{-1}$) & 0.005 & 0.64\\
\enddata
\tablenotetext{a}{The epoch corresponds to when the star reaches the superior conjunction in its orbit.  } 
\tablenotetext{$A$}{ The orbital elements for Gaia~Source~ID~454293469890174976}
\tablenotetext{$B$}{ The orbital elements for Gaia~Source~ID~3453566198941548160}
\end{deluxetable}

\section{Summary} \label{sec:summary}
Runaway stars are OB-type massive stars ejected from their birthplace through either the supernova ejection mechanism or the dynamical ejection scenario. 
The exact formation scenario is still a subject of debate. 
Motivated by the recent release of the large collection of spectroscopic observations for early-type stars from the LAMOST MRS DR8 and the unprecedented collection of proper motion measurements made by the \emph{Gaia} DR3, in this work, we report our search for the Galactic runaway stars through kinematic analysis for 4,432 OB-type stars. 
By following the procedures from \citet{1998Moffat,1999Moffat}, we calculated the peculiar velocities for the sample stars and determined a runaway criterion of $V_{\rm Sp} > 43$ km s$^{-1}$. 
Applying such a selection criterion results in identifying 229 runaway candidate stars from the sample. 

We investigated the statistical distribution of the identified runaway population based on their derived physical parameters from prior studies. 
Most of the runaway population are B-type stars (with $T_{\rm eff}$ values spanning the range of 10 to 20 kK).
A wide distribution in the projected rotational velocity ($V\sin{i}$) values of the runaway candidate stars is shown in our sample, with the finding that most of the runaway candidates are either rapidly rotating stars with small peculiar space velocity $V_{\rm Sp}$, slow spinning stars with large $V_{\rm Sp}$, or slow rotators with small peculiar space velocities. 
Only two stars with both high V sin i and VSp values were found.
We also determined the spatial location of the runaway sample and found that most of them likely reside within the Galactic thin disk. 
By inspecting the runaway population's RV variation, we noticed two runaway candidates display significant Doppler shifts. 
Preliminary single-lined circular orbital fits were applied to the RV measurements of the stars to determine their orbital solutions. 
These two binary candidate runaway stars have estimated orbital periods of 40 and 61 days.

This study has provided a new sample of Galactic runaway stars. 
Continuing monitoring of these targets and conducting further analysis to trace their past trajectories and find their parent clusters to determine their age will provide insights to help understand the formation scenario of the runaway stars.

\begin{acknowledgments}
This work is supported by the Natural Science Foundation of China (Nos.\ 12125303,12288102,12090040/3,12103064), the National Key R\&D Program of China (grant Nos. 2021YFA1600403, 2021YFA1600400) ,and by the China Manned Space Project of No. CMS-CSST-2021-A10. C.L.\ acknowledges the China Manned Space Project with no.\ CMS-CSST-2021-A08.
L.W.\ acknowledges the support by the NSFC under program No.\ of 12103085. 
This work is also supported by the ``Yunnan Revitalization Talent Support Program''—Science \& Technology Champion Project (NO.\ 202305AB350003), the Key Research Program of Frontier Sciences, CAS, Grant No.\ QYZDY-SSW-SLH007, and the International
Centre of Supernovae, Yunnan Key Laboratory (No.
202302AN360001).
This work is supported by the Natural Science Foundation of Yunnan Province (No. 202201BC070003).

Guoshoujing Telescope (the Large Sky Area Multi-Object Fiber Spectroscopic Telescope, LAMOST) is a National Major Scientific Project built by the Chinese Academy of Sciences. Funding for the project has been provided by the National Development and Reform Commission. LAMOST is operated and managed by the National Astronomical Observatories, Chinese Academy of Sciences.
\end{acknowledgments}

\bibliography{runaway}{}
\bibliographystyle{aasjournal}

\end{CJK*}
\end{document}